\documentclass[useAMS,usenatbib]{mn2e}
\usepackage[dvips]{graphicx}

\title[The Effect of Stellar Metallicity on the Sizes of Star Clusters]{The Effect of Stellar Metallicity on the Sizes of Star Clusters }
\author[Schulman, Glebbeek, and Sills]{Rafael D. Schulman$^{1,2}$, Evert Glebbeek$^{1,3}$, and Alison Sills$^{1}$\thanks{E-mail:
rschulma@uwo.ca (RDS); e.glebbeek@astro.ru.nl (EG); asills@mcmaster.ca
(AS)}\\
$^{1}$Department of Physics and Astronomy, McMaster University, Hamilton, ON L8S 4M1, Canada\\
$^{2}$Department of Physics and Astronomy, University of Western Ontario, London, ON, Canada\\
$^{3}$Department of Astrophysics, Radboud University Nijmegen, P.O. Box 9010 NL-6500 GL Nijmegen, The Netherlands\\}
\begin{document}

\maketitle

\label{firstpage}

\begin{abstract}
Observations indicate blue globular clusters have half-light
radii systematically larger than those of red globular clusters. In this paper, we test whether the different metallicity-dependent stellar evolution timescales 
and mass-loss rates within the clusters can impact their early dynamical
evolution. By means of N-body simulations including stellar evolution
recipes we simulate the early evolution of small centrally concentrated clusters with and without primordial mass segregation. Our simulations include accurate metallicity-dependent mass loss from massive stars. We find blue clusters to be larger than red clusters regardless of whether the clusters have been primordially mass segregated. In addition, the size difference is found to be larger and consistent with observations for initial models with a low central concentration. These results indicate that the systematic size difference found between red and blue clusters can, at least in part, be attributed to the dynamical effects of differing stellar evolution
histories, driven by metallicity.      
\end{abstract}


\section{Introduction}
Globular clusters are often used as probes of formation and evolution of their host galaxies. We use cluster properties such as mass, size, metallicity, and age to infer when and where the clusters formed, and therefore learn about the star formation history and merger history of the current hosts. However, these fundamental cluster properties are not completely understood in their own right. We know little about the initial conditions of cluster formation, and our understanding of the early evolution of star clusters is currently incomplete. In particular, the
early evolution of star clusters is complicated as it involves both stellar dynamics and
stellar evolution. Understanding the relevant processes driving the evolution is key to explaining many of the current observations
regarding the properties and distribution of globular clusters, such as the specific
frequency problem (see \citealp{Harrisrev} for a review). However, the theoretical
advancements in this field of study lag behind the observational
progress being made. In particular, making the leap from introducing
theoretical models which are merely special cases to
ones which are realistic and have a general applicability is an
on-going challenge. Thankfully, with the continuous
improvement of high-performance computation, the theoretical models
 are finally beginning to offer invaluable insight into the nature of globular clusters. 

The half-light radius, or effective radius ($r_{hl}$), of a globular cluster is a
measurable and theoretically useful quantity. There is increasing evidence that some size
and structural properties of globular clusters in cluster systems are intimately related
to the properties of the galaxy in which they are located
(e.g. \citealp{galaxy_struc1,galaxy_struc2}). Among these properties, the half-light radius is
particularly interesting because several models have shown that it
remains nearly constant throughout the lifetime of a globular cluster
\citep{rhlconstant1, rhlconstant2}. Therefore, this quantity allows
various globular clusters to be compared with little consideration of their exact
age. In this way, understanding the factors that determine the
half-light radii of clusters could provide helpful insight when
comparing the properties of different clusters. The half-light radius can provide
important constraints on the formation and subsequent evolution of
globular clusters. Another important observable property of clusters is their heavy-element abundance, or metallicity
($Z$). By studying the metallicities of globular clusters in a system, it is
possible to determine a metallicity distribution function for the
system. In doing so, a clear bimodal form emerges (see \citealp{Harrisrev} for a review). In this bimodal
distribution, there is a metal-poor mode centred at $Z \simeq
0.00063 $ and a metal-rich mode centred at $Z \simeq
0.0080$. Metal-poor and
metal-rich globular clusters are typically referred to as blue and red clusters respectively. 
Several studies (e.g. \citealp{puzia_obs, barmby_obs, Harris2009}) have studied the
half-light radii of red and blue clusters in elliptical galaxies, and
report a systematic size difference between the two. Specifically,
blue clusters are found to have a half-light radius $\sim20\%$ (or
$\sim0.5$ pc) larger than red
clusters in many galaxies such as M31 and NGC 4472. However, there is no consensus of the underlying reason
behind this discrepancy. Determining the explanation for this
observation could provide important insights into the formation and/or
evolution of globular clusters. 

\citet*{LarsenandBrodie} have proposed that this result can
be attributed purely to a projection effect. They argue that the
observed size difference can result from the differing
spatial distributions of the two cluster types in combination with a
correlation between galactocentric distance and cluster size. The argument
is supported by the observation that the size difference
becomes insignificant at large galactocentric distances
\citep{LarsenandBrodie}. However, \citet*{Harris2009}, in a more
detailed survey, reports the metal-poor clusters to be systematically
smaller by roughly the same factor at large galactocentric distances as well, favouring
the notion of a more intrinsic difference rather than merely a
projection effect. He goes on to offer another possibility for the
size discrepancy, which concentrates on
differences in formation conditions of the two globular cluster subpopulations. 
Specifically, he suggests that a higher
metallicity protocluster may undergo more rapid cooling and cloud
contraction before stars form, hence allowing a metal-rich cluster to
have a smaller scale size from the beginning. Yet another explanation
has been put forth by \citet{Jordan}, who claims the differences are
the result of the combined effects of mass segregation and the
dependence of stellar evolution time-scales on metallicity. The models
produced by \citet{Jordan} yield results consistent with
observations. However, in preparing the models, the important
assumption is made that the half-mass radii ($r_{hm}$) of different metallicity
clusters are equal. 

In this paper, we propose that the observed difference is indeed
linked to the dependence of stellar lifetimes and mass loss on
metallicity, and the impact this effect has on the dynamical evolution
of the cluster (similar to the explanation proposed by
\citet{Jordan}). We expand upon this idea by performing simple dynamical simulations, in which the metallicity-dependent stellar mass loss rates and stellar lifetimes are also included. In the early evolution of the clusters, there are chiefly two
mechanisms that drive the expansion: stellar mass loss and two-body
relaxation. In the first of these, stellar mass that is shed as the
stars evolve leaves the cluster and effectively reduces the total
cluster mass. The decreased mass results in outer stars feeling less
attraction to the core of the cluster, which in turn results in a bulk
expansion \citep{expansion}. In the second mechanism, when massive stars undergo close encounters with low-mass stars, the low-mass stars will
tend to be ejected from the encounter at high velocities, often
driving them to the outskirts of the cluster. This effectively causes
the cluster to expand \citep{expansion}. Although stellar mass loss boosts expansion by the first mechanism, it also opposes expansion by greatly reducing the
strength of two-body relaxation as the massive stars are lost. In this
sense, these two mechanisms are opposing processes. The process that
is dominant will drive the expansion. The two processes can also be thought of as
causing energy flow and production respectively
\citep{time_scales}. A central energy source is present, as a
consequence of active stellar mass-loss, and is in balance with the energy
flow outwards caused by two-body relaxation. \citet{time_scales}
suggests that this interplay results in no sharp transition between a
stellar evolution dominated phase and a dynamics dominated
phase. It is important to note, that high-mass metal-rich stars lose more mass
in stellar winds and also have reduced main-sequence lifetimes \citep{MUSEmath}. Thus,
at any instant in time during the cluster evolution, a
high-metallicity cluster will have shed more of its mass through
stellar evolution processes and contain fewer
high-mass stars than its low-metallicity counterpart. This leads us to
expect that if stellar mass-loss is the dominant expansion mechanism,
a metal-rich cluster will be more prone to expanding, whereas if
two-body relaxation is dominant, the opposite will be true. Thus, for
metal-poor clusters to become larger, we expect expansion due to two-body relaxation to
be dominant for the majority of the evolution. 
We explore this line of reasoning with an N-body code and
stellar evolution recipes through MUSE \citep{MUSE}. MUSE is a multiphysics, multiscale software environment for modelling astrophysical systems. It consists of a series of modules for stellar dynamics, stellar evolution, and stellar hydrodynamics, written in different languages and often based on commonly-available codes which were specifically written for one of those tasks. The stellar
evolution fits used in this research are functions of metallicity and
yield a unique mass loss history for each metallicity.  
We employ a modest number of stars, to test whether this effect can be important for the evolution of clusters. The impact of differential stellar evolution
and mass loss caused by variations in metallicity  
witnessed in these small clusters should, in general, be applicable to clusters with any number
of stars. 

\section{Method}
Simulating the early evolution of star clusters is a complex task, requiring
detailed treatment of both stellar evolution and gravitational
interactions. MUSE is
well suited for this problem as it combines stellar evolution and
dynamics modules into one package and allows calculations to be
carried over between modules with ease. We use NBODY-1h \citep{nbody1h}, a
direct N-body integrator, to follow the stellar dynamics. This early version of the NBODY series of dynamics codes provides the most consistent treatment of collisional stellar dynamics of the modules available in MUSE. NBODY-1h employs a softening parameter, $\epsilon$, in the gravitational force so that very close encounters between point masses do not dominate the computation. 
We choose the softening
parameter to be small, such that $\epsilon^{2} = 10^{-11}$ (in
dynamical units), which corresponds to $\epsilon = 0.65$ AU in our simulation.
We use single-star evolution (SSE) fitting formulae \citep{MUSEmath} to perform stellar evolution
calculations. SSE is continuously updated, and the version implemented into MUSE was available in early 2009 (Hurley 2009, personal communication).
SSE is composed of a series of analytical formulae that fit observational
stellar evolution tracks and accurately model the effect of
metallicity on the mass loss and stellar life-times of stars. In particular,
high-metallicity massive stars have shorter main-sequence lifetimes and a higher mass loss rate from stellar winds. We assume that any stellar mass that is shed leaves the cluster instantaneously. Additionally, we neglect hydrodynamics, radiative transfer, and collisions between stars. The initial stellar population does not contain any primordial binaries. Recent simulations of young star clusters are starting to include many of these effects \citep[e.g.][]{GoodwinBastian,PZMcMMak}. However, to test the effect of metallicity, we have chosen to be true to the history of stellar dynamics, and are looking at a simplified system first. 

The initial distribution of stars within the cluster is given by a
King model \citep{King66} in which the stars are assumed to be in virial equilibrium.
The models are generated using Starlab \citep{STARLAB}. To accentuate dynamical effects, we employ a high initial
central potential ($W_{0}=$12) for the majority of our models, but
also perform tests at $W_{0}=$6. We use a \citet{Kroupa}  mass function to generate an appropriate initial mass spectrum for the
cluster. All stars begin as zero-age main sequence stars. The computations are chiefly performed on clusters that contain
8192 stars initially. Further simulations are performed on
1024-, 4192-, and 16384-star clusters to test for convergence of
results. The tidal radius for these clusters is set to 20 pc, which is appropriate for open clusters of a few thousand stars in the Milky Way. Unbound stars outside the
tidal radius of the cluster are removed from the simulation. Removing
escapers affects the simulation in two ways: once several stars are
removed, the computational time decreases; and the half-mass radius is
effectively pushed inwards. 
Simulations are performed at 5 different metallicities: $Z =
0.02$, $0.00796$, $0.001$, $0.000632$, and $0.0001$. These values were chosen to span the range from solar metallicity ($Z=0.02$) to a typical globular cluster value ($Z=0.001$) to the most metal-poor clusters ($Z=0.0001$). The most important metallicities to investigate
are $Z = 0.00796$ and $0.000632$, as these correspond to the mean
metallicities of red and blue clusters respectively \citep{Harrisrev}.

To test the effects of primordial mass segregation on a
given cluster, it is first evolved dynamically for 1.2 half-mass relaxation times
($t_{rh}$), that is, without allowing the stars to undergo stellar evolution. The primordial dynamical evolution leaves the cluster mass
segregated at the time stellar evolution is
initiated. We choose this method for simulating primordial mass
segregation, as was done by \citet{mass_seg_method}, since it is a self-consistent and convenient method
of generating 
mass segregated clusters. However, there are alternate methods for simulating
primordial mass segregation (e.g \citealp{mass_seg_alt1,mass_seg_alt2}).  Several have suggested that mass segregation
occurs on a timescale on the order of a relaxation time
(e.g. \citealp{mass_seg}). A useful quantity to describe the
relaxation time of a cluster is the half-mass relaxation time. We investigated the strength of the mass
segregation for clusters primordially mass segregated between 0.2 and
1.5 $t_{rh}$. We found the majority of the segregation to occur within
the first 1.2 $t_{rh}$, that is, by this point in time, most massive
stars have migrated to the core of the cluster. Therefore, we chose this length of time as the
standard for all subsequent primordial mass segregation. We do not
believe that our method of simulating primordial mass segregation will
have a significant impact on the results that we are interested
in, because the differences between metallicities should only be
sensitive to whether or not the cluster has been primordially
segregated, and not to the method that has been used to achieve the
segregated state.  

Table \ref{table:simulation} lists all the simulations that have been performed along with
their specific information and alias. We use simulation e) as a
standard and will compare other runs to this particular one. It is
important to note that in all mass segregated models, the
central potential refers to that of the initial model. That is, the
specified $W_{0}$ applies to the initial model before mass segregation
has occurred. In fact, the central potential at the point of
initiation of stellar evolution is considerably lower. Similarly, the values of $N$ listed
for all primordially mass segregated models also apply to the initial
model. On the other
hand, all half-mass relaxation times shown are valid at the point of
initiation of stellar evolution, as well as the initial mass, $M_{0}$,
and the initial half-mass radius, $r_{h0}$ . Simulation m) is the result of an
average of three separate runs. This averaging is done to reduce the
statistical noise present at such a low number of particles. We evolve
the simulations for 100 Myr or 5 $t_{rh}$, whichever is
longer. By 100 Myr, stars more massive than $\sim 7 M_{\sun}$ have completed their
evolution and have become remnants, so the impact of any metallicity dependence will already be
present. 5 $t_{rh}$ is sufficiently long for the cluster to relax
dynamically. 

 \begin{table*}
 \centering
 \begin{minipage}{140mm}
  \caption{List of Simulations Performed}
  \begin{tabular}{@{}cccccccc@{}}
\hline
   Alias & $N$& $Z$  &Primordial&$W_{0}$&$t_{rh}$ [Myr]&
   $M_{0}$ [$M_{\sun}$]& $r_{h0}$ [pc] \\
&&&Mass Segregation&&&& \\

 \hline
 a) & 8192 & 0.02 & no& 12& 19.1& 14340.1& 0.94\\
b) & 8192 & 0.00796 & no& 12& 19.1& 14340.1& 0.94 \\
c) & 8192 & 0.001 & no& 12& 19.1& 14340.1& 0.94\\
d) & 8192 & 0.000632 & no& 12& 19.1& 14340.1& 0.94\\
e) & 8192 & 0.0001 & no& 12& 19.1& 14340.1& 0.94\\
f) & 8192 & 0.02 & yes& 12& 49.5& 13308.4& 1.77\\
g) & 8192 & 0.00796 & yes& 12& 49.5& 13308.4& 1.77\\
h) & 8192 & 0.001 & yes& 12& 49.5& 13308.4& 1.77\\
i) & 8192 & 0.000632 & yes& 12& 49.5& 13308.4& 1.77\\
j) & 8192 & 0.0001 & yes& 12& 49.5& 13308.4& 1.77\\
k) & 16384 & 0.0001 & no & 12 & 26.0& 26598.0& 0.96\\
l) & 4096 & 0.0001 & no & 12& 15.4& 6578.3& 0.94\\
m) & 1024 & 0.0001 & no & 12& 10.5& 1669.2& 0.96\\
n) & 8192 & 0.00796 & no & 6& 15.5& 13055.0& 0.80\\
o) & 8192 & 0.0000632 & no & 6& 15.5& 13055.0& 0.80\\

\hline
\end{tabular}
\end{minipage}
\label{table:simulation}
\end{table*}

\section{Results}
\begin{figure} 
 \includegraphics[width=0.5\textwidth]{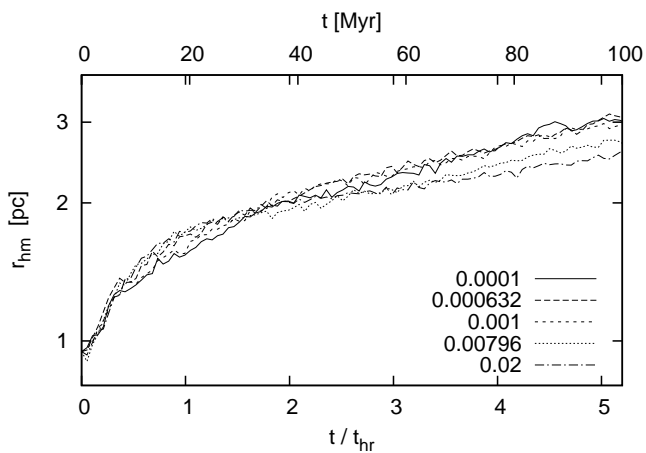}
 \caption{The half-mass radius of an 8192-star cluster
   evolved at various metallicities without primordial mass
   segregation. The clusters experience an initial rapid period 
   expansion and then continue to expand until the end of the
   evolution. The low metallicity clusters become systematically
   larger than the high metallicity clusters over the course of the evolution. }
   \label{fig:rh_evolution}
\end{figure}

In Figure \ref{fig:rh_evolution} we show the early size evolution of clusters a) to e). The
half-mass radius as a function of time is plotted vs both a dynamical and physical time-scale. All clusters experience a rapid expansion initially. The early evolution differs for metal-rich and metal-poor
clusters. In particular, over the course of the evolution, the
metal-poor clusters (c), d), and e)) become gradually larger over
time, in comparison to the metal-rich clusters. In the
evolution, the metal-rich clusters are seen to
expand more rapidly at first, but are soon overtaken by the metal-poor
clusters. Past this crossover point, the metal-poor clusters remain
larger. Although \citet{time_scales} indicates that there is no
sharp transition between a stellar dominated phase and a dynamics
dominated phase, in Figure \ref{fig:rh_evolution} we notice a gradual transition between the two. From this, we can conclude that in the early evolution (first
$\sim$20 Myr), stellar mass-loss is the important expansion mechanism,
but for the rest of the evolution, two-body relaxation dominates.  At the end of the simulation, the physical age of the cluster is
100 Myr. Thus, most of the stellar evolution of the massive stars will have been
completed at this point. Due to this, we expect effects of the
differing stellar evolutions between metallicities to have impacted
the dynamical evolution and be clearly evident. The fact that our results indicate the half-mass radius of 
the clusters to depend on
the metallicity contradicts the fundamental assumption made in
\citet{Jordan}. He assumes that the half-mass
radius of clusters is independent of metallicity to determine the
dependence of the observed half-light radius on metallicity. 
We calculated the half-light radii of a number of our clusters at 5 $t_{rh}$, and compared them to the half-mass radii. In these young stellar systems (unlike Jordan's 13 Gyr old populations), we do not see a systematic dependence of the half-light radii on metallicity, independent of the half-mass radius of the cluster. The ratio of half-mass to half-light radii was constant for all our clusters, within 10\%. 

\begin{figure} 
 \includegraphics[width=0.47\textwidth]{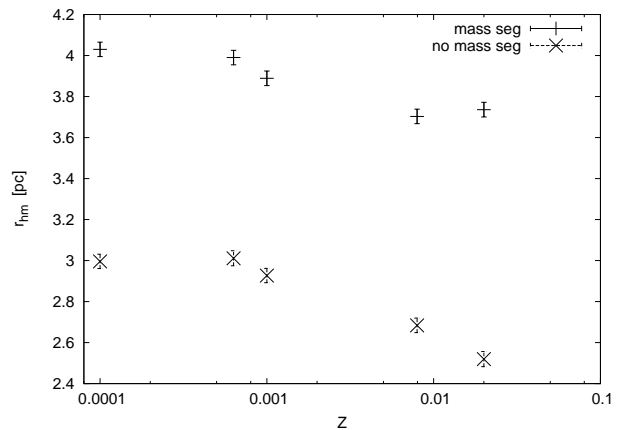}
 \caption{The  half-mass radius of an 8192-star cluster
    at 95 Myr at various metallicites with and  without primordial mass
   segregation. In both cases, the half-mass radius of the metal-poor
   clusters is larger. }
   \label{fig:rh_95Myr}
\end{figure}

\begin{figure} 
 \includegraphics[width=0.47\textwidth]{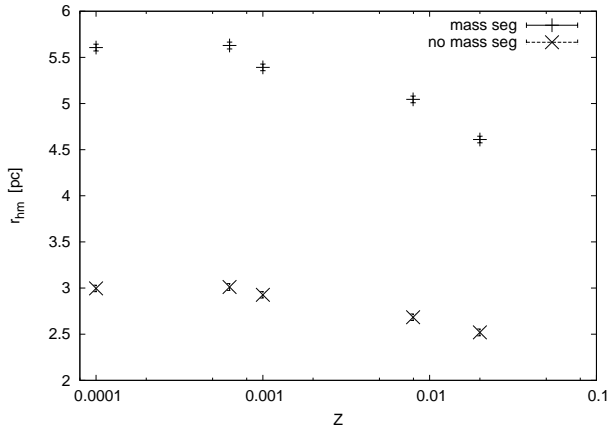}
 \caption{ The  half-mass radius of an 8192-star cluster
    at 5 $t_{rh}$ at various metallicities with and  without primordial mass
   segregation. Once again, the half-mass radius of the metal-poor
   clusters is larger, although the difference is more significant at
   this point than at 95 Myr for the primordially mass segregated models.  }
   \label{fig:rh_5trh}
\end{figure}

In Figure \ref{fig:rh_95Myr} and Figure \ref{fig:rh_5trh}, we plot the half-mass
radius as a function of metallicity at 95 Myr and 5 $t_{rh}$
respectively for runs a) through i). The value of the half-mass radius
plotted in these figures is the time-averaged quantity over 10 Myr
centred at 95 Myr and 5 $t_{rh}$ respectively. For the clusters
without primordial mass segregation (a) through e)), this corresponds to the same
instant in time. For the clusters that are primordially mass
segregated, 5 $t_{rh}$ corresponds to a physical age of roughly 250
Myr. The error bars are based on the range of points over the 10 Myr average. At 95 Myr we are comparing clusters of similar stellar evolution stage, while at 5 $t_{rh}$ we are comparing clusters of similar dynamical age.

\begin{figure}
\includegraphics[width=0.5\textwidth]{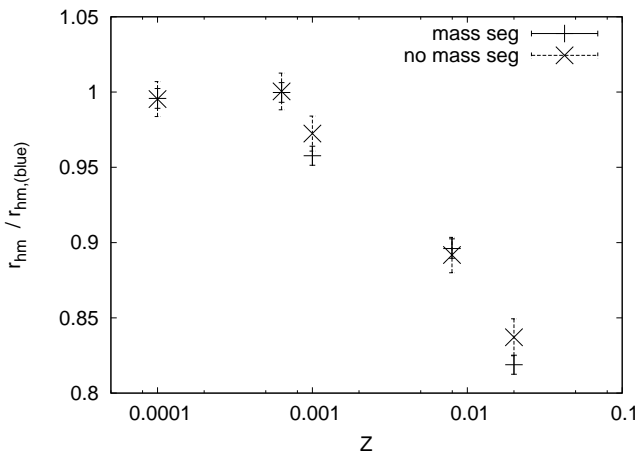}
\caption{The half-mass radius of an 8192-star cluster at 5 $t_{rh}$ at various metallicities, with and without mass segregation, normalized by the half-mass radius of the Z=0.00796 model. The trend of cluster size vs metallicity is independent of the amount of primordial mass segregation.}
\label{fig:rh_compare}
\end{figure}

Our results show that 
metal-rich clusters are smaller in size, regardless of whether the
cluster is primordially mass segregated. In fact, metal-poor clusters
of varying metallicity are very similar in size. The same property is
somewhat present in metal-rich clusters of varying metallicity as well. At 95 Myr (or 5
$t_{rh}$), the
difference in size between the red and blue metallicity clusters,
without primordial mass segregation, is 0.32$\pm$0.05 pc or (12$\pm$2) \%. With primordial mass segregation, the size
difference is 0.29$\pm$0.05 pc or (8$\pm$1)\% at 95 Myr. At 5
$t_{rh}$, the size difference between primordially mass segregated
blue and red metallicity clusters is instead 0.58$\pm$0.05 pc or (12$\pm$1)\%. In fact, at 5 $t_{rh}$ (Figure \ref{fig:rh_compare}), the trends with and without primordial mass segregation are very
similar. Thus, in the primordially mass segregated clusters as well, our results indicate a significant relationship between half-mass radius and metallicity.

There have been several observational reports of the size discrepancy between red
and blue clusters. A size difference ranging between 17\%$-$30\% has been found in many galaxies, including  NGC 4472 \citep{puzia_obs} and M31 \citep{barmby_obs}. More recently, \citet*{Harris2009} has reported a difference in
$r_{hl}$ of (17$\pm$2)\% at all galactocentric distances for the
globular cluster populations in six giant elliptical galaxies. In a
survey of 43 early-type galaxies, \citet{masters_obs} observe red clusters
and blue clusters to have a mean $r_{hl}$ of 2.8$\pm$0.3 pc and
3.4$\pm$0.3 pc respectively. This discrepancy corresponds to a
percent difference within the range listed previously. \citet*{Harris2009}
points out that the difference in $r_{hl}$ is typically $\sim$ 0.5$ - $1
pc. However, since the number of stars in the clusters we have
simulated are orders of magnitudes smaller, we believe that a
comparison of the percent difference in size is a more appropriate
diagnostic. On the other hand, we see that the observed physical size
difference is on par with that indicated by our results. It is
possible that this is due to that the results we gather for 8192-star
clusters are applicable to much larger clusters as well. That is to
say, the results converge for large $N$. In this way, our clusters do
indeed simulate real sized clusters quite accurately. Convergence of results will
be discussed later in this section. Although the percent differences we report fall short of the range that is observed, it is important to note that the clusters
we simulate are still expanding at the time that we compare their
sizes (See Figure \ref{fig:rh_evolution}). This implies that the size differences we report at these points may be subject to a small amount of change. As an example, the primordially mass segregated cluster shows a difference in the size
discrepancy when compared at 95 Myr as opposed to 5 $t_{rh} \simeq$ 250
Myr. However, we note that no major differences will develop past the
point of $\sim$100 Myr because all the massive stars have completed
their stellar evolution. Thus, the main impact of metallicity
differences should already be present in the cluster. Our results clearly indicate the
size of clusters to be metallicity dependent. In fact, it is this
qualitative agreement between our results and observations that is the most important. The quantitative differences could simply be a consequence of the parameter space we are occupying or the specifics of our simulations and
analysis. Although, as previously stated, we should be careful when
comparing the physical size differences to observations, it is interesting
to note that after 5 $t_{rh}$, the blue and red primordially mass
segregated clusters differ by roughly 0.6 pc at the half-mass
radius. This value is in the range of size differences that
\citet*{Harris2009} points out as relevant. 

\begin{figure} 
 \includegraphics[width=0.5\textwidth]{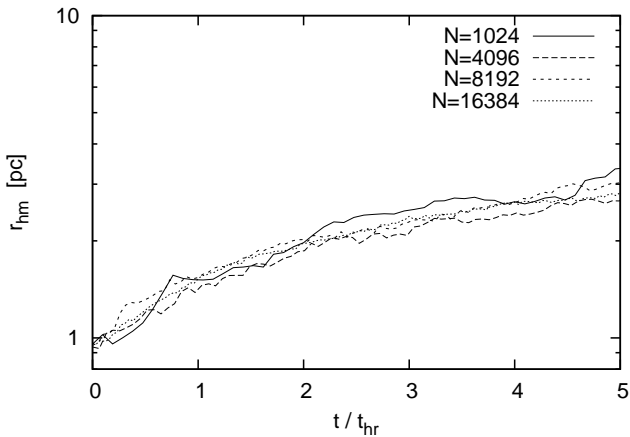}
 \caption{The  half-mass radius of clusters with varying
   star numbers, $N$, plotted on a dynamical time-scale. The evolution
   is dynamically similar at all $N$. }
   \label{fig:convergence}
\end{figure}

Figure \ref{fig:convergence} shows the results of runs e), k), l), and m) on a dynamical timescale, to illustrate the similarity of the behaviour of the half-mass radius regardless of the number of stars we use in our simulations. Although these are the lowest metallicity clusters in our sample, the behaviour of stellar dynamics is not strongly dependent on metallicity and therefore the convergence tests can be applied all our metallicities. The convergence of results at higher numbers of stars points to the fact that the results presented in this paper are applicable for a much larger number of stars. Indeed, although we study the metallicity effects in clusters of 8192 stars, the results
are applicable to all star clusters, with numbers of stars up to 2 orders of magnitude larger. This is in agreement with earlier studies, such as those of \citet{BaumMak}, which span a larger range in star number than we used here. The early evolution, and the fractional stellar mass loss from all size of clusters, is largely independent of the number of stars in the cluster.

Simulations n) and o) represent 8192-star clusters with a low initial
central potential ($W_{0}$=6) at red and blue metallicity
respectively. The tests were performed to investigate whether the size discrepancy remains if the
initial central potential is lower.  Indeed, at 5 $t_{rh} =$ 77.5 Myr,
the half-mass radius of the blue cluster is  0.39$\pm$0.05 pc larger, or
(21$\pm$3)\%. At 95 Myr, the difference is instead 0.41$\pm$0.05 pc ,
or (20$\pm$3)\%. These results are in the
range of the observed size differences. Therefore, we find a better
agreement between our results and observations when $W_{0}$ is low. 

\section{Summary \& Discussion}

In this paper, we have used an N-body code and stellar evolution recipes including
stellar mass loss to simulate
the early evolution of star clusters of different metallicities. In our
analysis, we include clusters with and without primordial mass
segregation. We focus our attention on the half-mass radius of the
clusters, as we find it to convey the same results as the half-light
radius. Clusters without primordial mass segregation demonstrate
roughly a (12$\pm$2)\% difference in half-mass radius between red and blue
clusters at 5 $t_{rh} =$ 95 Myr. Clusters with primordial mass
segregation exhibit an (8$\pm$1)\% difference in half-mass radius between red
and blue metallicities at 95 Myr and a (12$\pm$1)\% difference at 5
$t_{rh} \simeq$ 250 Myr. The simulation
was repeated at constant metallicity for $N=$ 1024, 4096, 8192, and 16384. On
a physical time-scale, the evolution converges at larger $N$,
indicating that our results are indeed applicable to much larger
clusters. On a dynamical time-scale, the evolution is very similar
regardless of $N$, which suggests that the dynamical evolution is
typically similar regardless of the number of stars in the
cluster. The size difference between blue and red clusters was found
not only to persist, but to be even more significant, at a low value
of the initial central potential. Therefore, we conclude that the
dependence of stellar evolution and mass loss histories on metallicity
impacts the dynamics, and at least in part, is responsible for the
size differences observed between red and blue globular clusters.

There is increasing evidence that globular clusters begin their evolution
primordially mass segregated. Although mass segregation occurs
naturally over the course of the evolution of a globular cluster, the process is
disturbed by the stellar evolution that is occurring simultaneously. Thus, a primordially mass
segregated cluster experiences a different dynamical
evolution. Therefore, for the sake of generality, we run simulations of both
primordially mass segregated and unsegregated models. Our results
demonstrate that the size difference persists in both cases.
In a primordially mass segregated cluster, the initial configuration
consists of the majority of massive stars clustered in the core
amongst a great number of low-mass stars, and additional low-mass
stars occupying the outer regions of the cluster. It is expected that
in this configuration, there will be an increase in the number of
close encounters between low- and high-mass stars. Due to this, and since we observe that two-body
relaxation is the dominant expansion mechanism for the majority of
the evolution in the unsegregated case, that should also be the case
for the primordially mass segregated case. In addition, due to the
increase in the number close encounters between low- and high-mass
stars, we anticipate a primordially mass segregated cluster to
produce even greater size discrepancies between metallicities. However, we
observe the size difference to be equal at 5 $t_{rh}$ and even larger
for the unsegregated cluster at 95 Myr.

The so-called specific frequency problem in globular cluster systems refers to the inconsistency between the 
number of metal-poor halo stars per unit globular cluster at the same metallicity
and the equivalent ratio for metal-rich halo stars. In fact, there are roughly five times
fewer metal-poor halo stars per unit globular cluster at the same metallicity (see
\citealp{Harrisrev} for a review). This discrepancy is inconsistent
with the notion that halo stars come from tidally stripped or disrupted globular clusters (if this were
the case, one would expect the two ratios to be equal). One solution to
this problem is that metal-rich clusters are more easily disrupted,
thus accounting for the inflated number of high metallicity halo
stars. If a cluster indeed is to be more easily disrupted, one would
expect it to expand more over the course of its evolution, and in
doing so, is more susceptible to tidal stripping and evaporation of
stars. However, our results indicate that it is metal-poor clusters
that experience a larger expansion over their evolution. Thus, our
results are incompatible with this solution to the specific frequency
problem.

Numerous assumptions and simplifications have been made to make the
problem more tractable. Most importantly, we
perform our tests on clusters that contain roughly 8000 stars, whereas
globular clusters typically have $10^{5}$-$10^{6}$ stars. The main concern here is
that the effects witnessed in our smaller clusters are no longer present
or relevant in larger clusters. Although our results indicate the
cluster evolution to converge at larger values of $N$, this does not guarantee that the convergence remains at
$N\sim10^{5}$. On the other hand, we believe that our analysis of
the processes causing the size discrepancy to be valid regardless of
the size of the cluster since 
metallicity differences will always alter the main sequence lifetimes of
massive stars. Because of this, large clusters of different
metallicity will have different number of massive stars at any point
in the early evolution. Metallicity will affect the impact 
of two-body relaxation, and thus significantly impact the subsequent
evolution. Therefore, we are confident that the qualitative
results of our simulations will persist at larger $N$, although the
exact value of the size difference may not be consistent.
 
We perform the bulk of our simulations on clusters with an extremely high initial central
potential. This was initially thought to accentuate dynamical effects, and in doing
so, would emphasize differences in the evolution of the different
metallicity clusters. As was earlier described, the simulations were
repeated once with $W_{0}$=6 at red and blue metallicities. In this
case, we observe the size discrepancy not only to persist but also
be enhanced. This result demonstrates that a high initial central
potential does not necessarily accentuate differences in the interplay
between stellar and dynamical evolution. Although a $W_{0}$ of 6 is considered a
fairly low value of the parameter, this result demonstrates that our results are applicable
at both low and high values of $W_{0}$. Therefore, we anticipate our
results to be relevant to clusters within this range and typical
values of globular clusters. A possible explanation for seeing an enhanced
effect at small $W_{0}$ could be attributed to the relative
time-scales of the dynamics and the stellar evolution. For the
mass-loss differences between metallicities to play the greatest role
in altering the dynamical evolution, the typical two-body relaxation
time-scale should be on the order of the stellar evolution time-scale.
In this regime, one would expect to see the greatest interplay between
the effects. Using the central relaxation time, $t_{c}$, as a measure of the
relevant dynamical time-scale, we postulate that the closer this
quantity is to the typical time-scale of the stellar evolution, the
greater the size discrepancies between different metallicity clusters
will be. In our clusters, the central relaxation times are
shorter than the stellar evolution time-scale of even the most massive stars. However, the $W_{0}=$6
cluster we use has a $t_{c}$ nearly 2 times larger than the
$W_{0}=$12 cluster. This implies that the relevant dynamical
time-scale for the $W_{0}$=6 cluster is more comparable to the
stellar evolution time-scale. Consequently, this cluster experiences
the greater difference between varying metallicities. According to the
2010 update of \citet{catalog}, the majority of Milky Way globular clusters that have not
experienced core collapse have $t_{c} \sim$ 30 Myr - 1000 Myr. Stars
of mass roughly between 2 and 10 $M_{\sun}$ have lifetimes in this
range. Thus, the dynamical time-scale is comparable to the stellar
evolution time-scale for stars in this range. Stars with even higher
masses will have stellar lifetimes shorter than 30 Myr, but
differences in their mass loss histories will still have an impact on the
dynamics of the cluster. By this, metallicity
differences are expected to alter the dynamical
evolution of globular clusters, and it is thus consistent with our analysis. 

In all our simulations, we assume a constant tidal radius of
20 pc. Although changing this parameter would alter the Lagrangian
radii (distances from the cluster centre at which various percentages
of the total mass are enclosed) over the course of the evolution, it would affect clusters of different metallicities in the same way, and thus would not change our
results. Many additional simplifications have been made to the system
including instantaneous ejection of gas, no primordial binaries, and no
binary stellar evolution. In general, ignoring these processes does
impact the evolution significantly. However, most of these processes
have been thoroughly studied in the past, and their effects are
thought to be well understood. We believe that we are justified in
neglecting these because they will not produce effects that are
significantly metallicity dependent. Thus, removing these
simplifications may alter the evolution of the cluster, but not in
ways that will change the size discrepancy between clusters of different metallicities.  

Our simulations have not included the effects of a population of primordial binaries in the cluster. Binaries are known to affect the dynamical evolution of clusters, mainly through the slowing or halting of core collapse. In the early dynamical stages of cluster evolution, as presented here, however, the effect of binaries is expected to be minimal. What has yet to be investigated in detail, however, is the effect of binarity on stellar mass loss and the subsequent implications for the dynamical evolution of clusters. Stellar mass loss in binary systems could well be enhanced, due to the presence of the companion; it could also be suppressed in some cases because the presence of a close companion means that the star does not reach a giant stage with enhanced mass loss but instead interacts and possibly merges with its companion. A more careful treatment of binary populations and binary evolution, especially for binaries of low metallicity, is required before we can fully understand the effects of binary populations on the sizes of red and blue globular clusters. 

\section{Acknowledgments}
RS is supported by the NSERC Undergraduate Student Research Award program. AS is supported by NSERC. This work was made possible by the facilities of the Shared Hierarchical Academic Research Computing Network (SHARCNET:www.sharcnet.ca) and Compute/Calcul Canada.  Many thanks also to Nathan Leigh, Jeremy Webb, and Rachel Ward for helpful discussions, comments, and insights. 
\label{lastpage}

\bibliography{mn}
\bibliographystyle{mn2e}

\end{document}